\documentclass[11pt,a4paper]{article}
\usepackage{amssymb,latexsym,a4wide}
\def\proof{{\bf Proof}.\ }
\def\bull{\vrule height .9ex width .8ex depth -.1ex }

\newtheorem{formula}{}[section]

\newtheorem{definition}[formula]{Definition}
\newtheorem{corollary}[formula]{Corollary}
\newtheorem{remark}[formula]{Remark}
\newtheorem{lemma}[formula]{Lemma}
\newtheorem{theorem}[formula]{Theorem}

\def\thrm{\begin{theorem}}
\def\thrml#1{\begin{theorem}\label{#1}}
\def\ethrm{\end{theorem}}
\def\rmrk{\begin{remark}}
\def\rmrkl#1{\begin{remark}\label{#1}}
\def\ermrk{\end{remark}}
\def\dfntn{\begin{definition}}
\def\dfntnl#1{\begin{definition}\label{#1}}
\def\edfntn{\end{definition}}
\def\nmrt{\begin{enumerate}}
\def\enmrt{\end{enumerate}}

\def\qtn{\begin{equation}}
\def\qtnl#1{\begin{equation}\label{#1}}
\def\eqtn{\end{equation}}
\def\lmm{\begin{lemma}}
\def\lmml#1{\begin{lemma}\label{#1}}
\def\elmm{\end{lemma}}
\def\crllr{\begin{corollary}}
\def\crllrl#1{\begin{corollary}\label{#1}}
\def\ecrllr{\end{corollary}}
\begin{document}
\title{Weak B\'ezout inequality for D-modules}
\author{
Dima Grigoriev \\[-1pt]
\small IRMAR, Universit\'e de Rennes \\[-3pt]
\small Beaulieu, 35042, Rennes, France\\[-3pt]
{\tt \small dima@math.univ-rennes1.fr}\\[-3pt]
\small http://name.math.univ-rennes1.fr/dimitri.grigoriev
}
\date{}
\maketitle
\begin{abstract}
Let $\{w_{i,j}\}_{1\leq i\leq n, 1\leq j\leq s} \subset
L_m=F(X_1,\dots,X_m)[{\partial
\over
\partial X_1},\dots, {\partial \over
\partial X_m}]$ be linear partial differential operators of orders with respect to
${\partial
\over
\partial X_1},\dots, {\partial \over
\partial X_m}$ at most $d$. We prove an upper bound
\begin{eqnarray}
n(4m^2d\min\{n,s\})^{4^{m-t-1}(2(m-t))} \nonumber
\end{eqnarray}
\noindent
on the leading coefficient of the Hilbert-Kolchin polynomial of the left
$L_m$-module
$\langle \{w_{1,j}, \dots , w_{n,j}\}_{1\leq j \leq s} \rangle \subset L_m^n$
having
the differential type $t$ (also being equal to the degree of the Hilbert-Kolchin
polynomial). The main technical tool is the complexity bound on solving systems of
linear
equations over {\it algebras of fractions} of the form
$$L_m(F[X_1,\dots , X_m, {\partial
\over
\partial X_1},\dots, {\partial \over
\partial X_k}])^{-1}.$$
\end{abstract}
\section*{Introduction}
Denote the derivatives $D_i={\partial
\over
\partial X_i}, 1\leq i \leq m$ and by $A_m=F[X_1,\dots , X_m, D_1, \dots , D_m]$ the
{\it Weil algebra} \cite{B} over an infinite field $F$. It is well-known that
$A_m$ is
defined by
the following
relations:
\qtnl{0}
X_iX_j=X_jX_i,D_iD_j=D_jD_i,X_iD_i=D_iX_i-1,X_iD_j=D_jX_i,\quad  i\neq j
\eqtn

For a family $\{w_{i,j}\}_{1\leq i\leq n, 1\leq j\leq s} \subset L_m$ of elements
of the
{\it algebra of linear partial differential operators} one can consider a system
\qtnl{-1}
\sum_{1\leq i \leq n} w_{i,j}u_i=0, \quad 1\leq j\leq s
\eqtn
\noindent
of linear partial differential equations in the unknowns $u_1,\dots , u_n$. In
particular,
if the $F$-linear space of solutions of (\ref{-1}) has a finite dimension $l$ then
the
quotient of the free $L_m$-module $L_m^n$ over the left $L_m$-module $L=
\langle \{w_{1,j}, \dots , w_{n,j}\}_{1\leq j \leq s} \rangle \subset L_m^n$
has also the dimension $l$ over the field $F(X_1,\dots , X_m)$ \cite{K}. Denote by
$t$
the {\it differential type} of $L$ \cite{K}, then $0\leq t \leq m$ (observe that
the case
treated in the previous sentence, corresponds to $t=0$).

We consider the filtration on the algebra $L_m$ defined on the monomials by $ord(
cD_1^{i_1}\cdots D_m^{i_m})=i_1+\cdots +i_m$ where a coefficient $c\in F(X_1,\dots
, X_m)$.
With respect to this filtration the module $L$ posesses the Hilbert-Kolchin
polynomial
\cite{K}
$${l \over t!}z^t+l_{t-1}z^{t-1}+\cdots +l_0$$
\noindent
of the degree $t$ (which coincides with the differential type of $L$). The leading
coefficient $l$ is called the {\it typical differential dimension} \cite{K}. In
the
treated above particular case $t=0$ the dimension of $F$-linear space of solutions
of
(\ref{-1}) equals to $l$.

In the present paper we prove (see Section~\ref{bezout}) the following inequality
which could be viewed as a weak analogue of the B\'ezout inequality for
differential
modules.
\crllrl{bez}
Let $ord(w_{i,j})\leq d, \quad 1\leq i \leq n, 1\leq j \leq s$. Then the leading
coefficient of the Hilbert-Kolchin polynomial
    \begin{eqnarray}
l\leq n(4m^2d\min\{n,s\})^{4^{m-t-1}(2(m-t))} \nonumber
    \end{eqnarray}
\ecrllr

Actually, one could slightly improve this estimate while making it more tedious.
We
note that the latter estimate becomes better with a smaller value of $m-t$. In
fact,
for
small values $m-t\leq 2$ much stronger estimates are known. In the case $m-t=0$
the
bound $l\leq n$ is evident. In the case $m-t=1$ the bound $l\leq \max_{1\leq i\leq
s}
\{ord(w_{i,1}\} +\cdots + \max_{1\leq i\leq s}
\{ord(w_{i,n}\}$ was proved \cite{K} (moreover, the latter bound holds in the more
general situation of  {\it non-linear} partial differential equations, whereas in
the
situation under consideration in the present paper of {\it linear} partial
differential equations  a
stronger {\it Jacobi conjecture} was
established , see e.g. \cite{P}). In the case
$m-t=2, \quad n=1$ the bound $l\leq ord(w_1)ord(w_2)$ was proved for the left
ideal
$\langle w_1, w_2, \dots \rangle \subset L_m$ where $ord(w_1)\geq ord(w_2) \geq
\dots$
\cite{P} which could be viewed as a direct analogue of the B\'ezout inequality. In
the
case $m=3, \quad t=0, \quad n=1$ a counter-example of a left ideal $\langle
w_1,w_2,w_3
\rangle \subset L_3$ is also produced in \cite{P} which shows that the expected
upper
bound
$ord(w_1)ord(w_2)ord(w_3)$ on $l$ appears to be wrong. It would be interesting to
clarify how sharp is the estimate in  Corollary~\ref{bez} for large values of
$m-t$.

The main technical tool in the proof of   Corollary~\ref{bez} is the complexity
bound
on solving linear systems over {\it algebras of fractions} of $L_m$. Let $K\subset
\{1,\dots , m\}$ be a certain subset. Denote by $A_m^{(K)}=F[X_1,\dots , X_m,
\{D_k\}_{k
\in K}] \subset A_m$ the corresponding subalgebra of $A_m$. We consider the
algebra of
fractions $Q_m^{(K)}=A_m(A_m^{(K)})^{-1}$. For an element $a\in A_m$ we denote the
Bernstein filtration \cite{B} $deg(a)$ defining it on monomials $X_1^{j_1} \cdots
X_m^{j_m} D_1^{i_1} \cdots D_m^{i_m}$ by $j_1+\cdots +j_m+i_1+\cdots +i_m$. Then
for an
element
  $ab^{-1}\in Q_m^{(K)}, \quad
a\in A_m, b\in A_m^{(K)}$ we write that the degree $deg(ab^{-1})\leq \max
\{deg(a), deg(b)\}$.

In Section~\ref{fraction} below we study the properties of $Q_m^{(K)}$ and the
complexity
bounds on manipulating in $Q_m^{(K)}$. In Section~\ref{matrice} we establish
complexity
bounds on {\it quasi-inverse} matrices over the algebra $Q_m^{(K)}$. Finally, in
Section~\ref{system} we consider the problem of solving a system of linear
equations over
the algebra $Q_m^{(K)}$:
\qtnl{1}
\sum_{1\leq i\leq p}a_{j,i}V_i=a_j, \quad 1\leq j \leq q
\eqtn
\noindent
where the coefficients $a_{j,i}, a_j \in A_m, \quad deg(a_{j,i}), deg(a_j) \leq
d$.
We prove the following theorem.
\thrml{solution}
If (\ref{1}) is solvable over $Q_m^{(K)}$ then (\ref{1}) has a solution with
\begin{eqnarray}
deg(v_i)\leq (16m^4d^2(\min\{p,q\})^2)^{4^{m-|K|}} \nonumber
\end{eqnarray}
\ethrm

Assume now that the ground field $F$ is represented in an efficient way, say as a
finitely generated extension either of $\mathbb Q$ or of a finite field (see e.g.
\cite{G86}). Then one can define the bit-size $M$ of the coefficients in $F$ of
the input
$\{a_{j,i}, a_j\}$.
\crllrl{time}
One can test the solvability of (\ref{1}) and if it is solvable then yield some
its
solution in time
polynomial in
\begin{eqnarray}
M,\enskip q, \enskip p^m, \enskip (md\min\{p,q\})^{4^{m-|K|}m} \nonumber
\end{eqnarray}
\ecrllr

Theorem~\ref{solution} and  Corollary~\ref{time} generalize the results from
\cite{G90} established for the algebra $Q_m^{(\emptyset)}=L_m$ of linear
differential
operators to the algebras of fractions $Q_m^{(K)}$. In \cite{G90} it is noticed
that
due to the example of \cite{M} the bounds in  Theorem~\ref{solution} and
Corollary~\ref{time} are close to sharp.

The problem in question generalizes the one of solving linear systems over the
algebra of
polynomials which was
studied in \cite{S} where the similar complexity bounds were proved.
Unfortunately, one
cannot extend directly the method from \cite{S} (which arises to G.Hermann) to the
({\it non-commutative}) algebra  $Q_m^{(K)}$ because the method involves the
determinants.
Nevertheless, we exploit the general approach of \cite{S}.

We mention also that certain algorithmical problems in the algebra of linear
partial
differential operators were posed in \cite{G}.
\section{Algebra of fractions of differential operators}
\label{fraction}
Let a matrix $B=(b_{i,j}), \quad 1\leq i\leq p-1, 1\leq j \leq p$ have its entries
$b_{i,j} \in A_m^{(K)}$ and $deg(b_{i,j})\leq d$. The following lemma was proved
in
\cite{G90}.
\lmml{vector}
There exists a vector $0\neq c=(c_1,\dots , c_p) \in (A_m^{(K)})^p$ such that
$Bc=0$
and moreover, $deg(c)\leq 2(m+|K|)(p-1)d=N$.
\elmm

\proof Consider an $F$-linear space $U\subset (A_m^{(K)})^p$ consisting of all the
vectors
$c=(c_1,\dots , c_p)$ such that $deg(c)\leq N$. Then $\dim U =p {N+m+|K| \choose
m+|K|}$.
For any vector $c\in U$ we have $deg(Bc)\leq N+d$, i.e. $Bc \in W$ where the
$F$-linear
space $W$ consists of all the vectors $w=(w_1,\dots , w_{p-1})\in
(A_m^{(K)})^{p-1}$ for
which $deg(w)\leq N+d$, thereby $\dim (W)=(p-1){N+d+m+|K| \choose m+|K|}$.

Let us verify an inequality $p {N+m+|K| \choose m+|K|} > (p-1){N+d+m+|K| \choose
m+|K|}$
whence lemma would follow immediately. Indeed,
$${N+d+m+|K| \choose m+|K|}/{N+m+|K| \choose m+|K|}={N+d+m+|K| \over N+m+|K|}\cdots
{N+d+1 \over N+1}\leq \Bigl( {N+d+1 \over N+1}\Bigr) ^{m+|K|}.$$
\noindent
It suffices to check the inequality $({N+d+1 \over N+1})^{m+|K|} < {p \over p-1}$.
The
latter
follows in its turn from the inequality
\begin{eqnarray}
(1+{1 \over p-1})^{1/(m+|K|)} > 1 + \Bigl({1\over m+|K|}\Bigr) {1 \over p-1} +
{1\over 2}\Bigl({1\over m+|K|}\Bigr) \Bigl({1\over m+|K|}-1\Bigr) {1\over (p-1)^2}
>
\nonumber \\
1+{1\over 2}\Bigl({1\over m+|K|}\Bigr) {1 \over p-1} > 1+{d \over N+1} \bull
\nonumber
\end{eqnarray}

Notice that Lemma~\ref{vector} implies that $A_m^{(K)}$ is an Ore domain \cite{B},
i.e.
the expressions of the form $b_1b_2^{-1}$ where $b_1,b_2 \in A_m^{(K)}$ constitute
an
algebra.
Below we use the following notations: letters $a,\alpha$ (respectively, $b,
\beta$) with
subscripts denote the
elements from $A_m$ (respectively, from $A_m^{(K)}$).
Our nearest purpose is to show that the expressions of the form $ab^{-1}$  also
constitute
an algebra $Q_m^{(K)}=A_m(A_m^{(K)})^{-1}$ (see
above the Introduction) and to provide complexity bounds on performing arithmetic
operations in $Q_m^{(K)}$.
To verify that the sum $a_1b_1^{-1}+a_2b_2^{-1}$ can be represented in the desired
form
$a_3b_3^{-1}$ we note first the following bound on a {\it (left) common multiple}
of
a family of elements from $A_m^{(K)}$ being a consequence of  Lemma~\ref{vector}.
\crllrl{multiple}
For a family $b_1,\dots , b_p \in A_m^{(K)}$ of the degrees $deg(b_1), \dots ,
deg(b_p) \leq d$ there exist $c_1, \dots , c_p \in A_m^{(K)}$ such that
$b_1c_1=\dots =b_pc_p \neq 0$ of the degrees $deg(c_1),\dots , deg(c_p) \leq
2(m+|K|)(p-1)d$.
\ecrllr

Evidently, the same bound holds also for a {\it right} common multiple of
$b_1,\dots , b_p$ which equals to $c_1^{'}b_1=\dots =c_p^{'}b_p$.

To complete the consideration of the sum one can find $c_1,c_2 \in A_m^{(K)}$ such
that
$b=b_1c_1=b_2c_2$ according to  Corollary~\ref{multiple}, then
$a_1b_1^{-1}+a_2b_2^{-1}=a_1c_1b^{-1}+a_2c_2b^{-1}=(a_1c_1+a_2c_2)b^{-1}$.

For an element $a\in A_m$ we denote by $ord^{(K)}(a)$ the filtration degree of $a$
with respect
to the symbols $\{ {\partial \over \partial X_j} \}$ for $j \not \in K$ and by
$deg^{(K)}(a)$
the filtration degree of $a$ with respect to the symbols
$X_1,\dots , X_m, \{ {\partial \over
\partial X_k} \}$ for $k\in K$.

Next we verify that $(A_m^{(K)})^{-1}A_m=A_m(A_m^{(K)})^{-1}$ relying on the
following
lemma.
\lmml{denominator}
Let $a\in A_m, \quad b\in A_m^{(K)}$ be such that $ deg^{(K)}(a), deg^{(K)}(b)\leq
d,
\quad
ord^{(K)}(a)=e$. Then there exist suitable elements $\alpha \in A_m, \quad \beta
\in
A_m^{(K)}$
such that $b\alpha=a\beta$ (or in other terms $\alpha \beta ^{-1}=b^{-1}a$) and
moreover,
$ord^{(K)}(\alpha)\leq ord^{(K)}(a), \quad deg^{(K)}(\alpha), deg^{(K)}(\beta)
\leq
2(m+|K|){e+m-|K| \choose e}d$.
\elmm

\proof Write down $\alpha= \sum_I D^I\beta_I$ where indeterminates $\beta_I \in
A_m^{(K)}$ and  the summation ranges over all the derivatives $D^I=\prod _{j\notin
K}
D_j^{i_j}$ with the orders $\sum _{j\notin K} i_j \leq e$. In a similar manner
$a=\sum _I D^Ib_I$. Then the equality $b\alpha=a\beta$ turns into a linear system
in
${e+m-|K| \choose e}$ equations in ${e+m-|K| \choose e}+1$ indeterminates $\beta,
\{ \beta _I \} _I$. Applying to this system Lemma~\ref{vector} we complete the
proof.
\bull


Lemma~\ref{denominator} entails that the product of two elements $a_1b_1^{-1}$ and
$a_2b_2^{-1}$ from $Q_m^{(K)}$ has again the similar form $a_3b_3^{-1}$: indeed,
let
$b_1^{-1}a_2=a_4b_4^{-1}$ for appropriate $a_4\in A_m, \quad b_4\in A_m^{(K)}$,
then
$a_1b_1^{-1}a_2b_2^{-1}=a_1a_4(b_2b_4)^{-1}$.

Finally, to complete the description of the algebra $Q_m^{(K)}$ we need to verify
that
the relation $\alpha \beta^{-1}=b^{-1}a\in Q_m^{(K)}$ being defined as
$b\alpha=a\beta$,
induces an equivalence relation on $Q_m^{(K)}$. To this end it suffices to show
that the
equalities $\alpha_1\beta_1^{-1}=b_1^{-1}a_1, \quad
b_1^{-1}a_1=\alpha_2\beta_2^{-1},
\quad \alpha_2\beta_2^{-1}=b_2^{-1}a_2$ imply the equality $\alpha_1\beta_1^{-1}=
b_2^{-1}a_2$. Due to Corollary~\ref{multiple} there exist $a_3,a_4\in A_m$ such
that
$a_3a_1=a_4a_2$, hence
$a_4b_2\alpha_2=a_4a_2\beta_2=a_3a_1\beta_2=a_3b_1\alpha_2$, therefore
$a_4b_2=a_3b_1$.
Because of that $a_4b_2\alpha_1=a_3b_1\alpha_1=a_3a_1\beta_1=a_4a_2\beta_1$, thus
$b_2\alpha_1=a_2\beta_1$ that was to be shown.

The following corollary summarizes the established above properties of the algebra
$Q_m^{(K)}$.
\crllrl{fraction+}
In the algebra of fractions $Q_m^{(K)}=A_m(A_m^{(K)})^{-1}=(A_m^{(K)})^{-1}A_m$
two
elements $a_1b_1^{-1}, a_2b_2^{-1} \in A_m(A_m^{(K)})^{-1}$ are equal if and only
if
there exists an element $\beta^{-1}\alpha \in (A_m^{(K)})^{-1}A_m$ such that
$\beta a_1=\alpha b_1, \quad \beta a_2=\alpha b_2$.
\ecrllr
\section{Quasi-inverse matrices over  algebras  of differential operators}
\label{matrice}
Let us call an $p\times p$ matrix $C=(c_{i,j})$  a
right (respectively, left) {\it quasi-inverse} to an $p\times p$ matrix
$B=(b_{i,j})$
where the entries $c_{i,j},b_{i,j}\in A_m^{(K)}$ if the matrix $BC$ (respectively,
$CB$)
has the diagonal form with non-zero diagonal entries. The following lemma was
proved in
\cite{G90}.
\lmml{quasi-inverse}
If an $p\times p$ matrix $B$ over $A_m^{(K)}$ has a right quasi-inverse (we assume
that
$deg(B)\leq d$) then $B$ has also a left quasi-inverse $C$ over $A_m^{(K)}$ such
that
$deg(C)\leq 2(m+|K|)(p-1)d$.
\elmm

\proof First observe that there does not exist a vector $0\neq b \in
(A_m^{(K)})^p$ for
which $bB=0$ since $A_m^{(K)}$ is a domain (see \cite{B} and also
Section~\ref{fraction}).
Consider the $p\times (p-1)$ matrix $B^{(i)}$ obtained from $B$ by deleting its
$i$-th
column, $1\leq i \leq
p$. Due to
Lemma~\ref{vector} there exists a vector $0\neq c^{(i)} \in (A_m^{(K)})^p$ such
that
$c^{(i)} B^{(i)} =0$ and $deg(c^{(i)})\leq 2(m+|K|)(p-1)d$. Then the $p\times p$
matrix
with the rows $c^{(i)}, \quad 1\leq i \leq
p$ is a left quasi-inverse of $B$. \bull


We note that a matrix $G$ over $A_m$ (or over $Q_m^{(K)}$) has a quasi-inverse if
and
only if $G$ is {\it non-singular}, i.e. has an inverse over the skew-field $
Q_m^{(\{1,\dots , m\})}=A_m(A_m)^{-1}$. The latter is equivalent to that $G$ has a
non-zero determinant of Dieudonn\'e \cite{A}. The rank $r=rk(G)$ is defined as the
maximal size of non-singular submatrices of $G$. The following lemma was proved in
\cite{G90}.
\lmml{block}
Let $G=(g_{i,j})$ be a $p_1\times p_2$ matrix over $A_m^{(K)}$ with the rank
$rk(G)=r$
and
assume that the $r\times r$ submatrix $G_1$ of $G$ in its left-upper corner is
non-singular. Let an $r\times r$ matrix $C_1$ over $A_m^{(K)}$ be a left
quasi-inverse
to
$G_1$.
Then one can find an $(p_1-r)\times r$ matrix $C_2$ over the algebra $Q_m^{(K)}$
such that
$$\left(\begin{array}{cc} C_1 0 \\ C_2 E \end{array} \right) G=
\left(
\begin{array}{ccc|c}
g_1& & 0 &\\
&\ddots & & *\\
0 & & g_r &\\
\hline
& \mathbf{0} & & \mathbf{0}\\
\end{array}
\right)
$$
\noindent
where $E$ denotes the unit matrix.
\elmm

\proof The matrix $C_2$ is determined uniquely by the requirement that in the
product of
matrices in the right-hand side the left-lower corner is zero. Then the
right-lower
corner
is zero as well by the definition of the rank.\bull

We proceed to solving  system~(\ref{1}).
Denote $r=rk(a_{j,i})$. After
renumerating the rows and columns one can suppose the $r\times r$ submatrix in the
left-upper corner of $(a_{j,i})$ to be non-singular. Applying
Lemma~\ref{quasi-inverse} to $r\times r$ submatrix $(a_{j,i}), \quad 1\leq i,j\leq
r$
one gets a matrix $C_1$, subsequently
applying Lemma~\ref{block} one gets a matrix $C_2$. If the vector $(C_2\enskip
E)(a_1,
\dots , a_q)$ does not vanish then system~(\ref{1}) has no solutions. Otherwise,
if
$(C_2\enskip E)(a_1,
\dots , a_q)=0$ then system~(\ref{1}) is equivalent to a linear system over
$Q_m^{(K)}$
of the following form (see Lemma~\ref{block}):
\qtnl{2}
g_jV_j+\sum_{r+1\leq i\leq p} g_{j,i}V_i=f_j, \quad 1\leq j \leq r
\eqtn
\noindent
where $g_j,g_{j,i},f_j \in A_m$.
Lemma~\ref{quasi-inverse} implies that
$deg(g_j),deg(g_{j,i}),deg(f_j)\leq (4m(r-1)+1)d$.
Fix for the time being a certain $i, \quad r+1\leq i \leq p$. Applying
Lemma~\ref{vector} to the $r\times
(r+1)$ submatrix which consists of the first $r$ columns and of the $i$-th column
of the matrix in the left-hand side of (\ref{2}), we obtain $h_1^{(i)},\dots ,
h_r^{(i)},h^{(i)} \in
A_m$ such that
\qtnl{3}
g_jh_j^{(i)}+g_{j,i}h^{(i)}=0, \quad 1\leq j \leq r
\eqtn
\noindent
Moreover, $deg(h_j^{(i)}), deg(h^{(i)})\leq 4mr(4m(r-1)+1)d \leq (16m^2r^2-1)d$.
\section{Complexity of solving a linear system over an algebra of fractions of
differential operators}\label{system}
In the present section we design an algorithm to solve a linear system (\ref{2})
over
$Q_m^{(K)}$.

Fix for the time being a certain $\gamma \notin K$. An arbitrary element $h\in
A_m$ can
be written  as
\qtnl{4}
h=\sum_{0\leq s \leq t} D_{\gamma}^s h_s=\sum_{S=\{s_{\delta}\}_{\delta \notin K}}
(\prod_{\delta \notin K} D_{\delta}^{s_{\delta}})h_S
\eqtn
\noindent
where $h_s\in A_m^{(\{1,\dots ,m\} \setminus \gamma)}, \quad h_S\in A_m^{(K)}$.
Denote the leading coefficient $lc_{\gamma}(h)=h_t\neq 0$. We say that $h$ is
{\it normalized with respect to $D_{\gamma}$} when $lc_{\gamma}(h) \in A_m^{(K)}$.
The
following lemma plays the role of the normalization for the algebra $Q_m^{(K)}$
(cf.
Lemma 2.3 \cite{Sit} or Lemma 4 \cite{G90}).
\lmml{normalization}
For any finite family $H=\{h\}\subset A_m$ there exists a non-singular $F$-linear
transformation of the $2(m-|K|)$-dimensional $F$-linear subspace of $A_m$ with the
basis
$\{X_{\delta}, {\partial \over \partial X_{\delta}}\}_{\delta \notin K}$ under
which the
vector $\{{\partial \over \partial X_{\delta}}\}_{\delta \notin K}$ is transformed
as
follows:
$$\{{\partial \over \partial X_{\delta}}\}_{\delta \notin K} \rightarrow
\Omega \{{\partial \over \partial X_{\delta}}\}_{\delta \notin K}$$
\noindent
where the $(m-|K|)\times (m-|K|)$ matrix $\Omega=(\omega_{\delta_1, \delta}),
\quad
\omega_{\delta_1, \delta} \in F$, and the vector
$$\{X_{\delta}\}_{\delta \notin K} \rightarrow (\Omega^T)^{-1}
\{X_{\delta}\}_{\delta \notin K}$$
\noindent
such that any transformed (under the transformation continued to $A_m$) element $\overline {h} \in A_m$ for $h\in H$ is
normalized
with respect to $D_{\gamma}$. Moreover, $deg_{D_{\gamma}}(\overline {h})=
ord^{(K)}(\overline {h})$.
\elmm

\proof One can verify that this linear transformation keeps the relations
(\ref{0}),
therefore, one can consider $A_m$ as a Weil algebra with respect to the
variables $\{X_k\}_{k\in K} \cup (\Omega^T)^{-1}
\{X_{\delta}\}_{\delta \notin K}$ and the corresponding differential operators
$\{ {\partial \over \partial X_k}\} \cup \Omega \{{\partial \over
\partial X_{\delta}}\}_{\delta \notin K}$
(cf. also \cite{G90}).

We rewrite (\ref{4}) as
$$h=\sum_{S_0=\{s_{\delta}\}_{\delta \notin K}}
(\prod_{\delta \notin K} D_{\delta}^{s_{\delta}})h_{S_0} + \Sigma_1$$
\noindent
where in the first sum all the terms from (\ref{4}) with the maximal value of the
sum
$\sum_{s_{\delta} \in S_0} s_{\delta} = ord^{(K)}(h)$ are gathered. Then the
leading
coefficient
$$lc_{\gamma}(\overline h)= \sum_{S_0}(\prod_{\delta \notin K}
\omega_{\gamma, \delta}^{s_{\delta}}){\overline h}_{S_0} \in A_m^{(K)}.$$
\noindent
Since the latter sum does not vanish if and only if the result of its linear
transformation
$$\sum_{S_0}(\prod_{\delta \notin K}
\omega_{\gamma, \delta}^{s_{\delta}}) h_{S_0} \in A_m^{(K)}$$
\noindent
with respect to $\Omega^T$ does not vanish as well, the set of the entries
$\{\omega_{\gamma, \delta}\}_{\delta \notin K}$ for which $lc_{\gamma}(\overline
h)$
does not vanish, is open in the Zariski topology (and thereby, is non-empty taking
into account that the ground field $F$ is infinite). Hence for an open set of the
entries $\{\omega_{\gamma, \delta}\}_{\delta \notin K}$ the leading coefficients
$lc_{\gamma}(\overline h)$ do not vanish for all $h \in H$. Therefore,
$deg_{D_{\gamma}}(\overline {h})=ord^{(K)}(h)=ord^{(K)}(\overline {h})$ and
thereby,
$\overline {h}$ is normalized with respect to $D_{\gamma}$.\bull

Applying Lemma~\ref{normalization} to the family $\{h^{(i)}\}_{r+1\leq i \leq p}$
constructed in (\ref{3}), we can assume without loss of generality that
$0\neq lc_{D_{\gamma}}(h^{(i)}) \in A_m^{(K)}, r+1\leq i\leq p$.

Consider a certain solution $v_i\in Q_m^{(K)}, \quad 1\leq i\leq p$ of system
(\ref{2}).
Fix some
$r+1\leq i\leq p$ for the time being. One can divide (from the right) $v_i$ by
$h^{(i)}$ with the remainder in $Q_m^{(K)}$ with respect to $D_{\gamma}$, i.e.
$v_i=h^{(i)}\phi_i+\psi_i$ for suitable $\phi_i,\psi_i \in Q_m^{(K)}$ such that
$deg_{D_{\gamma}}(\psi_i)< deg_{D_{\gamma}}(h^{(i)})=t$. Let $v_i=\sum_{0\leq
s\leq t_1}
D_{\gamma}^sv_{i,s}$ where $v_{i,s} \in A_m^{(\{1,\dots ,m\} \setminus \gamma)}$
and
$v_{i,t_1}=lc_{D_{\gamma}}(v_i)$. Taking into account that $h^{(i)}$ is normalized
with respect to $D_{\gamma}$, one can rewrite
$lc_{D_{\gamma}}(h^{(i)})D_{\gamma}^{t_1-t}
=D_{\gamma}^{t_1-t}lc_{D_{\gamma}}(h^{(i)})+\sum_{0\leq s\leq
t_1-t-1}D_{\gamma}^s\eta_s$
for appropriate $\eta_s \in A_m^{(K)}$. Thus, one can put the leading term of (the
quotient) $\phi_i$ to be
$\phi_{i,t_1-t}=D_{\gamma}^{t_1-t}(lc_{D_{\gamma}}(h^{(i)}))^{-1}lc_{D_{\gamma}}(v_i)
\in
Q_m^{(K)}$. Then $deg_{D_{\gamma}}(v_i-h^{(i)}\phi_{i,t_1-t})<t_1$ and one can
continue
the process of dividing with the remainder achieving finally  $\phi_i,\psi_i$.

For a fixed $1\leq j\leq r$ we multiply each of the equalities (\ref{3}) for
$r+1\leq i
\leq p$ from the right by $\phi_i$ and subtract it from the corresponding equality
(\ref{2}), as a result we get an equivalent to (\ref{2}) linear system
\qtnl{41}
g_j\psi_j+\sum_{r+1\leq i\leq p}g_{j,i}\psi_i=f_j, \quad 1\leq j\leq r
\eqtn
\noindent
for certain $\psi_j \in Q_m^{(K)}$. Since $deg(f_j),deg(g_{j,i})\leq (4m(r-1)-1)d,
\quad deg_{D_{\gamma}}(\psi_i)< deg_{D_{\gamma}}(h^{(i)})\leq (16m^2r^2-1)d$
(see the end of  Section~\ref{matrice}) we conclude that
$deg_{D_{\gamma}}(\psi_j) \leq N_1 \leq 16m^2r^2d, \quad 1\leq j \leq r$.

Represent $\psi_j=\sum_{0\leq s \leq N_1}D_{\gamma}^s\psi_{j,s}, \quad 1\leq j
\leq p$
for appropriate $\psi_{j,s} \in A_m^{(\{1,\dots ,m\} \setminus
\gamma)}(A_m^{(K)})^{-1}$.

For each $0\leq s\leq N_1$ we have
\qtnl{6}
g_jD_{\gamma}^s=\sum_{0\leq l\leq N_0}D_{\gamma}^lg_{j,s,l}^{(1)}, \quad
g_{j,i}D_{\gamma}^s=\sum_{0\leq l\leq N_0}D_{\gamma}^lg_{j,i,s,l}^{(1)}
\eqtn
\noindent
for appropriate $g_{j,s,l}^{(1)},g_{j,i,s,l}^{(1)} \in
A_m^{(\{1,\dots ,m\} \setminus \gamma)}$ where $N_0,deg(g_{j,s,l}^{(1)}),
deg(g_{j,i,s,l}^{(1)})\leq 16m^2r^2d$. Substituting the expressions (\ref{6}) in
(\ref{41}) and subsequently equating the coefficients at the same powers of
$D_{\gamma}$,
we obtain the following linear system over
$A_m^{(\{1,\dots ,m\} \setminus \gamma)}(A_m^{(K)})^{-1}$:
\qtnl{7}
\sum_{j,s}g_{j,s,l}^{(2)}\psi_{j,s}=g_l^{(2)}
\eqtn
\noindent
being equivalent to system~(\ref{41}) and thereby, to system~(\ref{1}), in other
words,
these systems are solvable simultaneously. Moreover, $g_{j,s,l}^{(2)},g_l^{(2)}
\in
A_m^{(\{1,\dots ,m\} \setminus \gamma)}, \quad deg(g_{j,s,l}^{(2)}),
deg(g_l^{(2)})\leq
16m^2r^2d$,
the number of
the equations in system~(\ref{7})  does  not exceed $16m^2r^2d$ and
the number of the indeterminates $\psi_{j,s}$ is less than $16pm^2r^2d$.

We summarize the proved above in this section in the following lemma.
\lmml{elimination}
A linear system~(\ref{1}) of $q$ equations in $p$ indeterminates  with the degrees
of
the coefficients $a_{j,i},a_j$ at most $d$ is solvable
over the algebra
$Q_m{(K)}$ if and only if the linear
system~(\ref{7})  is solvable over the algebra $A_m^{(\{1,\dots ,m\} \setminus
\gamma)}(A_m^{(K)})^{-1}$. System~(\ref{7}) in at most $16pm^2r^2d$
indeterminates and in at most $16m^2r^2d$ equations has the
coefficients from the algebra $A_m^{(\{1,\dots ,m\} \setminus \gamma)}$ of the
degrees
less than $16m^2r^2d$ where $r\leq \min\{p,q\}$ is the rank of the
system~(\ref{1}).

Moreover, if system~(\ref{7}) has a solution with the degrees not exceeding a
certain
$\lambda$ then system~(\ref{1}) has a solution with the degrees not exceeding
$\lambda+16m^2r^2d$.
\elmm

Thus, we have eliminated the symbol $D_{\gamma}$. Continuing by recursion applying
Lemma~\ref{elimination} we eliminate consecutively $D_{\delta}$ for all $\delta
\notin
K$ and finally yield a linear system
\qtnl{8}
\sum_{1\leq l \leq N_3}g_{s,l}^{(0)}V_l^{(0)}=g_s^{(0)}, \quad 1\leq s \leq N_2
\eqtn
\noindent
over the skew-field $A_m^{(K)}(A_m^{(K)})^{-1}$ with the coefficients
$g_{s,l}^{(0)},
g_s^{(0)} \in A_m^{(K)}$ where $N_2, deg(g_{s,l}^{(0)}), deg(g_s^{(0)}) \leq N_4=
(2m)^{4^{m-|K|}}(dr)^{3^{m-|K|}}$ and the number of the indeterminates $N_3\leq
pN_4$.
Notice that system~(\ref{8}) is solvable simultaneously with system~(\ref{1}).

As in Section~\ref{matrice} one can reduce (with the help of Lemma~\ref{block})
system~(\ref{8}) to the diagonal-trapezium form similar to (\ref{2}) with the
coefficients from the algebra $A_m^{(K)}$ having the degrees less than
$2(m+|K|)N_4^2$
due to Lemma~\ref{quasi-inverse}. Therefore, if system~(\ref{8}) has a solution in
the skew-field  $A_m^{(K)}(A_m^{(K)})^{-1}$ it should have a solution of the form
$v_l^{(0)}=(b_l^{(1)})^{-1}b_l^{(2)} \in
(A_m^{(K)})^{-1}A_m^{(K)}$ with the degrees $deg(b_l^{(1)}), deg(b_l^{(2)})\leq
2(m+|K|)N_4^2$ taking into account the achieved diagonal-trapezium form. Applying
Corollary~\ref{multiple} to $v_l^{(0)}$ one can represent $v_l^{(0)}=v_l^{(3)}
(v_l^{(4)})^{-1}$ for suitable $v_l^{(3)},v_l^{(4)} \in A_m^{(K)}$ with the
degrees
$deg(v_l^{(3)}), deg(v_l^{(4)})\leq 4(m+|K|)^2N_4^2, \quad 1\leq l \leq N_3$.
Hence
due to  Lemma~\ref{elimination} it provides a solution of system~(\ref{1}) over
the
algebra $Q_m{(K)}=A_m(A_m^{(K)})^{-1}$ with the
bounds on the degrees $N_5=4(m+|K|)N_4^2$. This completes the proof of
Theorem~\ref{solution}. \bull

Finally we observe that if system~(\ref{1}) has a solution it has also a solution
of
the form $v_i=c_ib^{-1}$ for appropriate $c_i\in A_m, \quad b\in A_m^{(K)}$ with
the
degrees $deg(c_i), deg(b)\leq (2(m+|K|)p+1)N_5, \quad 1\leq i \leq q$ due to
Corollary~\ref{multiple}. The
algorithm looks for a solution of system~(\ref{1}) just in this form with the
indeterminate coefficients over the field $F$ at the monomials in the symbols
$X_1,\dots , X_m, D_1, \dots , D_m$ and treat  (\ref{1}) or equivalently,
$\sum_{1\leq i\leq p}a_{j,i}c_i=a_jb, \quad 1\leq j \leq q$ as a linear system
over
$F$ in the indeterminate coefficients. This completes the proof of
Corollary~\ref{time}. \bull
\section{A bound on the leading coefficient of the Hilbert-Kolchin polynomial of a
linear differential module}\label{bezout}
In the sequel we use the notations from the Introduction.
If the degree $0\leq t\leq m$ of the Hilbert-Kolchin polynomial of the left
$L_m$-module $L$ equals to $m$ then the leading coefficient $l$ is at most $n$
\cite{K}.

From now on assume that $t<m$. For each $1\leq i_0 \leq n$ and any family $K=
\{k_0,\dots , k_t\} \subset \{1,\dots , n\}$ of $t+1$ integers there exists an
element
$0\neq (0,\dots ,0, b_{i_0}^{(0)},0,\dots ,0)\in L$ with a single non-zero
coordinate
at the $i_0$-th place where $b_{i_0}^{(0)}\in A_m^{(K)}(F[X_1,\dots , X_m])^{-1}$,
taking into account that the differential type of $L$ equals to $t$
(cf. Proposition 2.4 \cite{Sit}). Rewriting the latter condition as a system of
linear equations
$$\sum_{1\leq j\leq s} C_jw_{i,j}=0, \quad i\neq i_0, \quad \sum_{1\leq j\leq s}
C_jw_{i_0,j}=1$$
\noindent
in the indeterminates $C_1,\dots , C_s$ over the algebra $Q_m^{(K)}$ and making
use of
Theorem~\ref{solution} one can find a solution of this system in the form
$c_1=(b_{i_0})^{-1}a_{1,i_0},\dots ,
c_s=(b_{i_0})^{-1}a_{s,i_0}\in Q_m^{(K)}$ for suitable
$b_{i_0}\in A_m^{(K)}
, \quad a_{1,i_0},\dots ,a_{s,i_0} \in A_m$ with the degrees
$deg(b_{i_0}),deg(a_{1,i_0}),\dots , deg(a_{s,i_0}) \leq
(16m^4d^2(\min\{n,s\})^2)^{4^{m-t-1}}$. Thus, $0\neq (0,\dots ,0,
b_{i_0},0,\dots ,0)\in L$.

Applying Lemma~\ref{normalization} to the family $\{b_{i_0}\}_{1\leq i_0 \leq n}$
we
conclude that after an appropriate $F$-linear transformation $\Omega$ of the
subspace
with the basis
$D_{k_0},\dots , D_{k_t}$ and the corresponding transformation $(\Omega^T)^{-1}$
of
the subspace with the basis $X_{k_0},\dots , X_{k_t}$, one can suppose that
$b_{i_0}=\alpha_eD_{k_0}^e+\beta_{e-1}D_{k_0}^{e-1}+\cdots +\beta_0$ is normalized
with respect to $D_{k_0}$ where $0\neq \alpha_e \in F[X_1,\dots , X_m]$ and
$\beta_{e-1}, \dots , \beta_0 \in A_m^{(K\setminus \{k_0\})}$. The Hilbert-Kolchin
polynomial does not change under the $F$-linear transformation $\Omega$. Taking
into
account that these
transformations keep the relations (\ref{0}) of the Weil algebra (see the proof of
Lemma~\ref{normalization}), in the applications of these transformations below
we may preserve the same notations for the basis of the
resulting Weil algebra after transformations.

First we apply the described above construction to the family $K=\{1,\dots ,
t+1\}$
and obtain normalized elements $(0,\dots ,0,
b_{i_0}^{(1)},0,\dots ,0)\in L, \quad 1\leq i_0 \leq n$ with respect to $D_1$.
Thereupon consecutively we take $K=\{2,\dots , t+2\}, \dots , K=\{m-t,\dots , m\}$
and obtain elements $(0,\dots ,0,
b_{i_0}^{(2)},0,\dots ,0), \dots , (0,\dots ,0,
b_{i_0}^{(m-t)},0,\dots ,0) \in L, \quad 1\leq i_0 \leq n$ being normalized with
respect to $D_2, \dots , D_{m-t}$, correspondingly.

Hence any element in the quotient $F(X_1,\dots , X_m)$-vector space $L_m^n$ over
the
left $L_m$-module $L$ can be reduced to the form
$(\sum_Ih_{1,I}D_1^{i_1}\cdots D_m^{i_m}, \dots ,
\sum_Ih_{n,I}D_1^{i_1}\cdots D_m^{i_m})$ where the coefficients $h_{j,I}\in
F(X_1,\dots , X_m)$ and $i_1,\dots , i_{m-t}\leq
(16m^4d^2(\min\{n,s\})^2)^{4^{m-t-1}}$. This completes the proof of
Corollary~\ref{bez}. \bull

\vspace*{0.7 cm} {\bf Acknowledgements.}
The paper was done partially under the support of the Humboldt-Preis. The author is
grateful to Michel Granger, Fritz Schwarz, Serguey Tsarev for their attention
and to SCAI (Fraunhofer Institut) for the hospitality during the stay there.

\end{document}